\documentclass[aps,prx,twocolumn,amsmath,nofootinbib,longbibliography,amssymb,superscriptaddress,10pt]{revtex4-2}

\usepackage{amsmath}
\usepackage[english]{babel}
\usepackage{graphicx}
\usepackage[normalem]{ulem}
\usepackage{soul}
\usepackage{float} %added by Evyatar
\usepackage[svgnames]{xcolor} %added by Evyatar
\usepackage[colorlinks=true, linkcolor=Red , citecolor= NavyBlue,urlcolor = NavyBlue, breaklinks=true]{hyperref}
\usepackage{verbatim}
\usepackage{esint}
\usepackage{marginnote}
\usepackage{graphicx,mathtools,bm,bbm}
\graphicspath{
    {figures/}
}

\renewcommand{\vec}[1]{\boldsymbol{#1}}
\newcommand{\RNum}[1]{\uppercase\expandafter{\romannumeral #1\relax}}

\def \beq {\begin{eqnarray}}
\def \eeq {\end{eqnarray}}

\begin{document}

%\title{A criterion for strangeness in disordered metals}

%\title{Signatures of strangeness in the Lorentz ratio of non-Fermi liquids}

%\title{A Criterion for Strangeness in the Lorentz ratio of non-Fermi liquids}

%\title{A Criterion for Strange Metallicity in the Lorenz Ratio}

%\title{Inhomogeneity-Induced Time Reversal Symmetry Breaking and Circulating Currents in Twisted Bilayer Cuprate Superconductors}

\title{Inhomogeneity-Induced Time-Reversal Symmetry Breaking 
in Cuprate Twist-Junctions}

\author{Andrew C. Yuan}
\thanks{These authors contributed equally to this work.}
\affiliation{Department of Physics, Stanford University, Stanford, CA 94305, USA}

\author{Yaar Vituri}
\thanks{These authors contributed equally to this work.}
\affiliation{Department of Condensed Matter Physics, Weizmann Institute of Science, Rehovot 76100, Israel}

\author{Erez Berg}
\affiliation{Department of Condensed Matter Physics, Weizmann Institute of Science, Rehovot 76100, Israel}
\author{Boris Spivak}
\affiliation{Department of Physics, University of Washington, Seattle, USA}
\author{Steven A. Kivelson}
\affiliation{Department of Physics, Stanford University, Stanford, CA 94305, USA}
\affiliation{Rudolf Peierls Centre for Theoretical Physics, University of Oxford, Oxford OX1 3PU, United Kingdom}

\date{\today}

\begin{abstract}
%By symmetry, t
The lowest order Josephson coupling, $J_1(\theta)\cos(\phi)$, between two d-wave superconductors with phase-difference $\phi$ across the junction vanishes when their relative orientation is rotated by $\theta=\pi/4$. However, in the presence of inhomogeneity, $J_{1}({\bm r})$ is non-zero locally, with a sign that fluctuates in space. We show that such a random $J_1$ generates a global second-harmonic Josephson coupling, $J_2\cos(2\phi)$, with a sign that favors %a phase difference across the junction, 
$\phi = \pm \pi/2$, i.e., spontaneous breaking of time reversal symmetry. 
The magnitude of $J_2$ is substantially enhanced if the spatial correlations of $J_1(\bm{r})$ extend over large distances, such as would be expected in the presence of  large amplitude twist-angle angle disorder or significant local electronic nematicity.
We argue that this effect likely accounts for the recent observations in twisted Josephson junctions between high temperature superconductors. 
\end{abstract}

\maketitle

%\section{Introduction}

Twisting and stacking two-dimensional quantum materials has proven to be an extremely powerful tool, both in creating new interesting materials~\cite{cao2018correlated,cao2018unconventional,yankowitz2019tuning}
% \cite{TBG,TwistedTMDs}
and in probing their properties~\cite{polshyn_large_2019,inbar2023quantum}. %\cite{QTM,...}
In particular, it has long been proposed that measuring the Josephson coupling between two $d$-wave superconductors as a function of the twist angle, $\theta$, between them can reveal the pairing symmetry. Early experiments in the cuprate high temperature superconductors \cite{li1999bi} did not observe the predicted angle dependence of the Josephson coupling, in apparent contradiction to the known $d$-wave order parameter symmetry. 
In a 
notable set of recent experiments \cite{zhao2021emergent}, a $|\cos(2\theta)|$ dependence was finally observed, possibly resolving this puzzle.  However, still more recent  experiments \cite{XueTwist} have not reproduced this result. 
%reproduced the early results - with no strong twist-angle dependence of the critical current.  
The source of these discrepancies remains an open question. 

Here, we assume that there is %some 
an extrinsic explanation for this apparent non $d$-wave behavior, and focus on the experiments \cite{zhao2021emergent} that show the expected angle dependence.  In particular, for $\theta = %45^\circ
\pi/4$,  the lowest-order Josephson coupling $J_1$  vanishes by symmetry. %Interestingly, i
It was recently predicted \cite{can2021high} that 
the second-order Josephson coupling, $J_2$ (corresponding to an inter-plane coherent tunneling of two Cooper pairs) favors a spontaneously broken time reversal symmetry (TRS) state, where the phase difference between the two superconductors is $\pm \pi/2$. Indeed, the experiment in Ref. \cite{zhao2021emergent} found evidence that at $\theta=%45^\circ
\pi/4$, there is a substantial second-order Josephson coupling, revealed by measuring additional half-integer Shapiro steps. The precise microscopic mechanism of this second-order coupling remains to be clarified. 
%Experimentally, the sign of the second-order Josephson coupling has not yet been determined, and direct evidence for time reversal symmetry breaking is still lacking. 
Also still to be explored experimentally is the theoretical proposal that this could serve as a platform to realize chiral topological superconductivity \cite{Volkov2020,can2021high,tummuru2022josephson,mercado2022high,Margalit2022}, although the gap of the resulting state may be very small~\cite{Song2022}.

However, as we will discuss, the extreme anisotropy of Bi$_2$Sr$_2$CaCu$_2$O$_{8+\delta}$ (Bi-2212) -- the cuprate superconductor used in the twist-junction experiments -- poses a significant quantitative 
difficulty with the intrinsic mechanism for generating $J_2$. 
%As we will discuss, 
Specifically, an estimate based on measured quantities in bulk crystals implies an intrinsic value of $J_2$ that is too small to explain the experiments. 

\begin{figure}[t]
	\begin{center}
		\includegraphics[width=0.99\columnwidth]{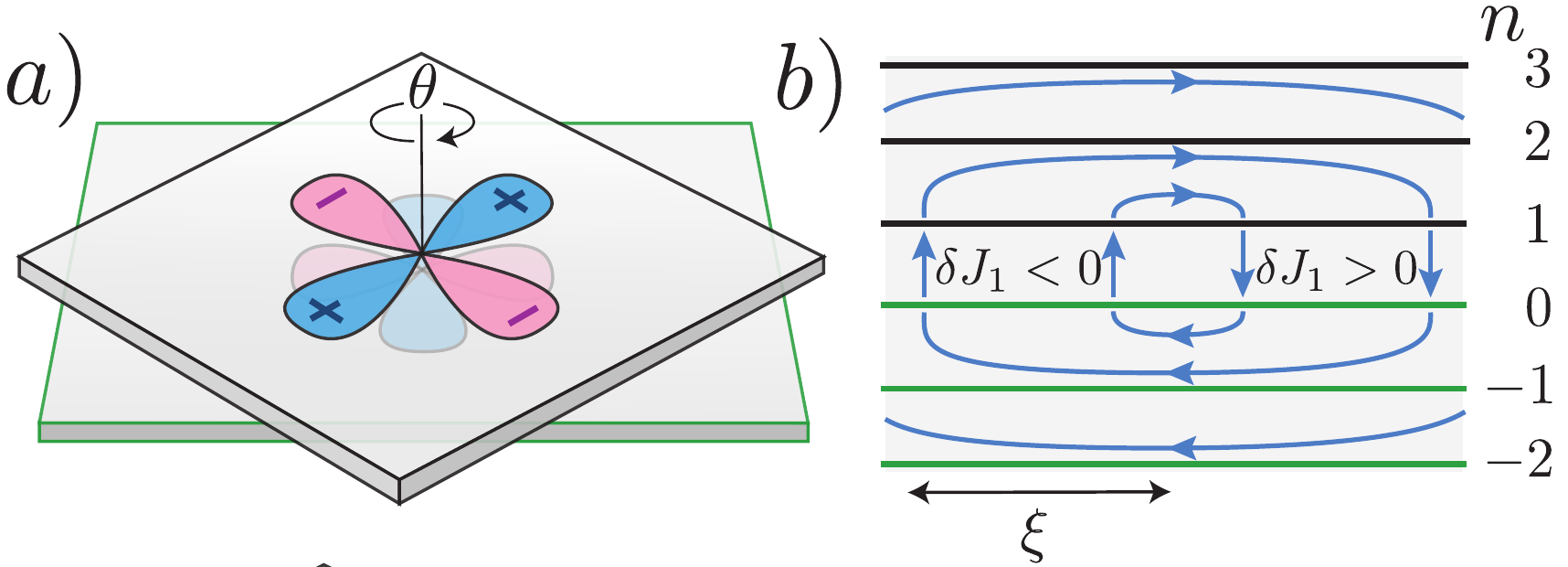}
	\end{center}
\vspace{-0.5cm}
    \caption{ a) A Josephson junction made of two $d$--wave superconductors with a relative twist angle $\theta$. For $\theta=\pi/4$, the lowest-order Josephson coupling $J_1$ vanishes by symmetry. b) Cross section of the junction region. The layers of the two superconductors are labeled by $n$. The twist junction is between the $n=0$ and $n=1$ layers. The Josephson coupling between these layers is given by Eq.~\ref{eq:flictuatingJ_{1}}, and its spatial average vanishes at $\theta=\pi/4$. The Josephson coupling $\delta J_1(\bm{r})$ is correlated over a length $\xi$. The ground state breaks time reversal symmetry spontaneously, and is characterized by circulating persistent current loops,  illustrated by blue arrows. 
	}
	\label{fig1}
\end{figure}

In this paper, we propose an alternative mechanism for the generation of $J_2$, a form of ``order from disorder''
driven by the effect of local spatial symmetry breaking.
In Fig.~\ref{fig1}a we schematically show a sample consisting of two $d$-wave superconductors
rotated with respect to each other by an angle $\theta$. We consider the case where each superconductor contains $N$ layers, and we label them by the index $n$ which extends from $-N+1$ to $N$.
We neglect the effect of inhomogeneity on all layers but those with $n=0,1$ (which are separated by the twist junction),  
and assume that the Josephson coupling per unit area between these layers has the form
\begin{equation}\label{eq:flictuatingJ_{1}}
J_{1}=\overline{J_{1}(\theta)} +\delta J_{1}({\bm r}) 
\end{equation}
Here $\overline{J_{1}(\theta)}$ respects the $d$--wave symmetry %and vanished 
so that it vanishes at $\theta=\pi/4$, while the random in sign quantity $\delta J_{1}$ is generated by a local, sample--specific point-group-symmetry breaking, such that $\overline{\delta J_1(\bm{r})} = 0$.
 Here the overline denotes configuration averages over a random ensemble of $\delta J_{1}({\bf r})$.

 The fluctuating component $\delta J_1(\bm{r})$ may arise from different physical sources. 
One possible source 
is spatial variations of the twist angle, $\delta \theta({\bm r})$, or any other form of disorder (including potential disorder) that breaks the local point group symmetries. 
A local admixture of an $s$--wave gap could also arise from 
``electron nematic domains.'' %of the type observed in scanning tunneling microscopy (STM) experiments in BSCCO~\cite{}. 
We will comment on these possibilities further in the discussion section below.  

We will characterise the fluctuations of $J_1$ by a correlation function
\begin{equation}
\overline{\delta J_1(\bm{r})\delta J_1(\bm{r}')}  = \overline {\delta J_1({\bm r})^2} 
\,f\left(\frac{|\bm{r}-\bm{r}'|}{\xi}\right)
\end{equation}
 Here, $f(\bm{x})$ is a dimensionless function normalized such that $f(\bm{0})=1$ and $\int d^2 x f(\bm{x})=1$, and $\xi$ is the correlation length of $\delta J_1(\bm{r})$. 
 %The subscript $s$ refers to the idea that $\delta J_{1}({\bm r})$ is related to the random in sign $s$--wave component of the order parameter, generated by a violation of the crystal symmetry.
 
The local inter-layer superconducting phase difference between the $n=0$ and $n=1$ layers, $\phi({\bm r})=\phi_1(\bm{r})-\phi_0(\bm{r})$, exhibits spatial fluctuations 
\begin{equation}
\phi({\bm r})=\bar{\phi} +\delta \phi({\bm r})
\end{equation}
which are correlated with the fluctuations of  the critical current density $\delta J_{1}({\bm r})$.
It is easy to show that at  $\theta = \pi/4$ the ground state of the system corresponds to $\bar{\phi}=\pm \pi/2$. 
Indeed, for $\bar{\phi} = \pm \pi/2$, %at small $\delta \phi({\bm r})$ 
the system gains energy linearly in $\delta \phi$ by adjusting the phase difference across the junction to the sign of the local Josephson coupling $\delta J_1(\bm{r})$. The energy cost of the resulting in-plane currents is only quadratic in $\delta \phi$. This simple argument also implies the violation of the time-reversal symmetry, and existence of {\it local} circulating supercurrents and associated magnetic fields, illustrated in Fig.~\ref{fig1}b. 
 Time reversal symmetry is broken in an interval of 
 twist angles near $\theta=\pi/4$. 
 
%Another aspect of the problem is that in typical situation 
Importantly, in the cuprates, the phase stiffness is strongly anisotropic, with the c-axis stiffness much smaller than the in-plane one. 
%rigidity $\kappa$ is much larger than $J_{1}$.  
As a result, there is a characteristic length in the problem (treating, for simplicity, the case of $N=1$ -- the case of $N\gg 1$ is discussed below)
\begin{equation}
\label{eq:l}
    \ell^2=
    \frac{\kappa}{
    (\overline{\delta J_1^{2}})^{1/2}},
    % \sim  \xi^{2}_{0}
    % \frac{J_{1}}{(\overline{\delta J_1^{2}})^{1/2}}\frac{\kappa_\parallel}{\kappa_\perp}, %,\,\,\,\ N=1
\end{equation}
% which separates two regimes. Here, 
where $\kappa$ is the in-plane phase stiffness.  
 %$\xi_{sc}$ is the coherence length of superconductor. 
 There are two regimes depending on $\xi/\ell$:
 In the case $\xi\gg \ell$ the system breaks into large domains with $\phi \approx 0$ or $\pi$,  separated by by domain walls. Time reversal symmetry is broken in the vicinity of the domain walls where the phase twists from $0$ to $\pi$. In this case, the critical current through  the junction is of the order of $(\overline{\delta J_1^2})^{1/2}$.  
 % the time-reversal symmetry is still broken, 
 % but the amplitude of the half-integer Shapiro steps  sequence of Shapiro steps in the I-V characteristics is small compared to the conventional one. 
 Below we show that in the opposite limit, 
 $\xi_\text{sc}\ll \xi \ll \ell$ (where $\xi_\text{sc}$ is the superconducting coherence length), 
 the energy--phase relation has the form
\begin{equation}\label{eq:effectingenergy}
    E(\bar{\phi}) = -\overline{J_1(\theta)} \cos(\bar{\phi}) - J_2\cos(2\bar{\phi}),
\end{equation}
 % In this case
 % the amplitudes of conventional and unconventional sequences of Shapiro steps are camparable, while, 
 where $J_{2}$ increases with $\xi$ (See Eqs.~\ref{eq:J2_res} below). 
  For $\xi\sim \ell$, we obtain 
%  on the boundary of thus interval we  arrive to a conclusion that this model can produce as much value of the critical current as  
\begin{equation}\label{eq:J2max}
 |J_2| \sim |J_{2}^{\rm{max}}|\sim 
\sqrt{{\overline{\delta J_{1}^{2}}}}. 
\end{equation}
Thus, our mechanism is capable of producing a $J_2$ that is as large as the magnitude of the local (random) Josephson coupling, 
given that the correlation length of the local Josephson coupling is sufficiently large.

In the presence of an in-plane magnetic field $B_\parallel$, another length scale arises: $l_B = \Phi_0/(B_\parallel w)$, where $w$ is the smaller of the perpendicular penetration depth and the thickness of the device, and $\Phi_0$ is the superconducting flux quantum. As long as $l_B$ is larger than $\xi$, we expect the system to behave essentially as a Josephson junction with a spatially uniform $J_1$ and $J_2$. In particular, at $\theta=\pi/4$, the expected Fraunhofer pattern %should 
will reflect the doubled periodicity of the energy-phase relation. At higher fields (such that $l_B<\xi$) a more complicated interference pattern will directly reflect the inhomogeneity of the Josephson coupling~\cite{Vituri}. 
 % From this prospective we are particularly motivated  by a host of experimental observations in BSCCO in which such inhomogeneities with large correlation length  have been directly imaged,(``nematic'') inhomogeneities.  Ref.\cite{...}

 % We also note that the outlined above scenario is not confined to the case of twisted layers of D-wave superconductors.  For example, the same analysis can be applied to weakly coupled S and D -wave superconductors. Finally, we mention that in the presence of finite amount of disorder the considered above electronic state is gapless, which, in particular means that it is not topological. 

{\it Estimate of the intrinsic $J_{2}$.--}
% in the second order in the interlayer tunneling amplitude.} %\cite{....}. }
%Below we calculate 
As a first exercise, we estimate the intrinsic second-harmonic Josephson coupling $J_{2}$ generated by coherent tunneling of two Cooper pairs, neglecting the effects of inhomogeneity, and argue that the intrinsic $J_2$ is too small to account for the experiment of Ref. \cite{zhao2021emergent}. 
% inhomogeneity-induced time-reversal symmetry breaking .

The measured values of the in-plane and out-of-plane penetration depths for optimally doped Bi-2212 are $\lambda_\parallel \sim 200$nm and $\lambda_\perp \sim 100\mu$m, respectively \cite{cooper1990direct}, corresponding to an anisotropic superfluid density, $\kappa_\perp/\kappa_\parallel=(\lambda_\parallel/\lambda_\perp)^2 \sim 4\times  10^{-6}$.  
This large anistropy reflects the extremely 2D electronic structure.  
To estimate the consequences of this for the expected ratio of $J_2/J_1$, we assume that the inter-layer tunneling is not momentum conserving (the ``incoherent tunneling model'' \cite{klemm2005phase}), which reproduces the approximately $\cos(2\theta)$ dependence of $J_1$ observed in the experiments .

For $\theta \sim 0$, $J_{1}(0)$ (which has units of energy per unit area) is determined from an appropriate average of the inter-layer hopping, $t_\perp$, by an Ambagoakar-Baratoff-like relation
\begin{align}
    J_1(0)= g_1 |t_\perp|^2 \frac{\nu^2}{k_F^2} |\Delta|, 
\end{align}
where $\nu$ is the density of states at the Fermi energy, $k_F$ is the Fermi wavevector, $|\Delta|$ is the size of the maximal gap, and $g_1$ is a  dimensionless factor that depends, among other things, on the structure of the tunneling matrix elements in momentum space~\cite{Bille2001,klemm2005phase}.  Similar reasoning gives for the intrinsic second order coupling
\begin{align}
J_2= g_2 |t_{\perp}|^{4}\frac{\nu^{3}}{k_{F}^{4}},
\label{eq:J2_BCS}
\end{align}
where $g_2$ is another dimensionless factor, and presumably $J_2$ is  approximately independent of $\theta$. 
%In Eq.~\ref{eq:J2_BCS}, we have assumed that the inter-layer tunneling is nearly incoherent~\cite{Bille2001}. 

To relate these $J_1$ and $J_2$ to measured quantities, we assume that the coupling between planes in a bulk crystal is the same as for the twist junction with $\theta\sim 0$, so that we can identify $\kappa_\perp = J_1(0) d$,  
where $d$ is the inter-bilayer distance. (For further consistency checks regarding the estimate of $J_1$ and $J_2$, see Supplementary Information.)  Moreover, the in-plane superfluid stiffness empirically satisfies $\kappa_\parallel =g_\parallel T_c/d$ where $g_\parallel$ is of order one.  Therefore
\begin{align}
    \left(\frac{J_2}{J_1(0)}\right)= \left(\frac{g_\parallel g_2}{g_1^2}\right) 
    \left(\frac{\xi_\text{sc}}{k_{F}d^{2}}\right)\left(\frac{T_{c}}{\left|\Delta\right|}\right)\left(\frac{\kappa_{\perp}}{\kappa_{\parallel}}\right).
    %\left(\frac{T_c}{|\Delta|}\right)\left(\frac{1}{k_F^2 d^2}\right)\left(\frac{\kappa_\perp}{\kappa_\parallel}\right). 
    %\sim 10^{-7},
 \label{estimate}
\end{align}
%where in making the final estimate, we have (without  justification) set the first factor (ratios of $g_j$'s) to 1, and taken experimental values for the remaining quantities.  
 Taking into account the fact that in optimally doped cuprates $T_c$ and $|\Delta|$ are of the same order of magnitude, and so are $1/k_F$, $d$, and $\xi_\text{sc}$, and ignoring the unknown numerical prefactors, we get that the intrinsic value of $J_2/J_1$ should be of the order of $10^{-6}-10^{-5}$. In contrast, the measured value of $J_2/J_1$ reported in Ref. \cite{zhao2021emergent} is $\sim 10^{-2}$. Thus, unless for some reason  $g_1$ is anomalously small, or $g_2$ unexpectedly large, the intrinsic effect is too small to account for the  cuprate experiment.

{\it Inhomogeniety--induced $J_{2}$.--} We use the following model to describe two layered superconductors coupled by a twisted Josephson junction: 
\begin{equation}
\label{eq:H}
\mathcal{H}=\sum_{n}\int d^2r\,\Big[\frac{\kappa}{2}\left(\nabla\phi_{n}\right)^{2}-\mathcal{J}^{(n)}\cos\left(\phi_{n}-\phi_{n-1}\right)\Big],
\end{equation}
where $\phi_n(\bm{r})$ is the superconducting phase in layer $n$ %and coordinate
at position $\bm{r}=(x,y)$, and $\kappa = \kappa_{\parallel}d$ is the phase stiffness in the plane. Each superconductor contains $N$ layers, and the sum over $n$ extends from $-N+1$ to $N$. For $n\ne 1$, the inter-plane coupling per unit area is $\mathcal{J}^{(n)} = J_{1}(0)$, related to the three-dimensional phase stiffness by $J_{1}(0) = \kappa_\perp/d \equiv J_\perp$. For $n=1$, the Josephson coupling is $\mathcal{J}^{(1)} = J_1(\bm{r})$, given by Eq. \ref{eq:flictuatingJ_{1}}.

% Note that the twist angle enters entirely through $h(\theta)$, whose form is determined by the $d$-wave symmetry of the order parameters of the two layers \cite{can2021high}. 

Note that we ignore magnetic screening throughout the analysis, neglecting the coupling of the superconducting phase to the electromagnetic field. This assumption is justified as long as the thickness of the system is much smaller than $\lambda_{\perp}$, and the lateral size is smaller than the Josephson penetration depth of the twist junction.  
We analyze the  system at $T=0$, minimizing the energy~\eqref{eq:H}. A similar analysis was performed in 3D case in Ref.~\cite{zyuzin2000theory}; here, we generalize the analysis for the case of layered superconductors and a finite correlation length of the local Josephson coupling $\delta J_1(\bm{r})$. 

For $\xi\ll \ell$ (where $\ell$ is defined in Eq. \ref{eq:l} for $N=1$ and in Eq. \ref{eq:ell_3D} below for $N\rightarrow \infty$), the superconducting phase is nearly uniform within each of the two superconductors. We therefore write the superconducting phase as $\phi_n(\bm{r}) = s_n \bar{\phi}/2 + \delta \phi_n(\bm{r})$, where $s_n=1$ for $n>0$ and $-1$ otherwise ($\bar{\phi}$ is the average phase difference), and expand the energy up to second order in $\delta \phi_n(\bm{r})$. Minimizing the resulting expression, we obtain the energy-phase relation of the junction per unit area given by Eq.~\ref{eq:effectingenergy} 
with the second-order Josephson coupling is given by
\begin{equation}
\label{eq:J2}
J_2 = -\int \frac{d^2 q}
{(2\pi)^2} \frac{\left|\delta J_{1}(\bm{q})\right|^{2}}
{\kappa q^{2}+J_{\perp}\frac{1-e^{-\eta_{\bm{q}}}+\alpha_{\bm{q}}\left(1-e^{\eta_{\bm{q}}}\right)}{1+\alpha_{\bm{q}}}}.
\end{equation}
(See Supplementary Information for further details.) Here, $\delta J_{1}(\bm{q})$ is the Fourier transform of $\delta J_1(\bm{r})$, $\eta_{\bm{q}} = \cosh^{-1}[1+\kappa q^2/(2J_\perp)]$, and 
\begin{equation}
\label{eq:alphaq}
\alpha_{\bm{q}} = e^{-2\eta_{\bm{q}}\left(N-1\right)}\frac{J_{\perp}\left(e^{\eta_{\bm{q}}}-1\right)-\kappa q^{2}}{J_{\perp}\left(1-e^{-\eta_{\bm{q}}}\right)+\kappa q^{2}}.     
\end{equation}
Crucially, the sign of $J_2$ is \emph{negative}. 
%(this follows from the fact that the integrand in Eq.~\ref{eq:J2} is positive). 
Therefore, at twist angle $\theta=\pi/4$ [such that $J_1(\theta)=0$], the ground state corresponds to $\phi = \pm \pi/2$. Time reversal symmetry is broken in a range of twist angles around $\pi/4$, such that $|J_2| > |J_1(\theta)|/4$. 
We provide explicit expressions for $J_2$ in two cases: 
\begin{equation}
\label{eq:J2_res}
    J_{2}\approx-\frac{\overline{\delta J_1^{2}}\xi^{2}}{2\pi\kappa}
    % J_{2}\approx-\frac{\kappa }{2\pi \ell^2} \frac{\xi^2}{\ell^2}
    \begin{cases}
        \ln\left(1+\frac{1}{\xi}\sqrt{\frac{\kappa}{J_{1}(0)}}\right), & N\rightarrow\infty,\\
        \ln\left(\frac{\kappa}{\left(\overline{\delta J^{2}}\right)^{1/2}\xi^{2}}\right), & N=1.
        % \ln\left(\ell^2/\xi^{2}\right), & N=1.
    \end{cases}
\end{equation}
The expressions are valid if $\xi$ is sufficiently short: $\xi\ll \ell$, such that $\delta \phi_n(\bm{r})\ll \pi$.
Beyond this regime  ($\xi>\ell$),  the perturbative calculation breaks down (see discussion around Eq. \ref{eq:l}).
%\addAY{(We may wish to change the order of the previous sentence and the following couple sentences, since $\ell$ hasn't been properly defined just yet.)} 
{The length scale $\ell$ can be identified as the value of $\xi$ where $J_2\sim (\overline{\delta J^2_1})^{1/2}$. Using Eq. \ref{eq:J2_res}, this gives  Eq.~\ref{eq:l} for $N=1$ (up to the logarithmic factor, which is of the order of unity in this case), whereas for $N\rightarrow \infty$ we find}
\begin{equation}
 \ell^2 \sim \frac{\kappa J_1(0)}{\overline{\delta J_1^2}}. 
 \label{eq:ell_3D}
\end{equation}
 
For $N=1$, the derivation of Eq.~\ref{eq:J2_res} contains a subtlety - the integral 
in Eq. \ref{eq:J2} is infrared divergent. This logarithmic divergence occurs also for any finite $N$, and signals a breakdown of the perturbative treatment in $\delta J_1$. A more accurate treatment (see Supplementary Information) reveals that the divergence is cut off by an emergent length scale\footnote{To be exact, a length scale of the order $\ell^2/\xi \times [\ln (\ell/\xi)]^{-1/2}$; see SI.}, of the order of $\ell^2/\xi$, which leads to Eq.~\ref{eq:J2_res}. 

% Substituting $\xi$ in Eq.~\ref{eq:J2_res} for $l$ we get the discussed above estimate for the maximum value of $J_{2}$, given by Eq.~\ref{eq:J2max}.

% It follows from Eq.~\ref{eq:J2_res} that the magnitude of the suppression of the Josephson coupling 
% at $\theta=45^{\circ}$ , $J_{2}/J_{1}(0)$, is quite sensitive to both the value of $\xi$ and to the amplitude of fluctuations of $\delta J_{1}$. From this prospective it is not surprising that different experiments, Refs.~\cite{zhao2021emergent, XueTwist} showed an order of magnitude difference in the amplitude of the suppression.

%In general, depending of the strength of disorder the amplitude of fluctuations of $\delta J_{1}$, and $J_{2}^{(max)}$ can be even comparable to $J_{1}$.
%Therefore we believe that described above mechanism can  explain the experiments in BSCCO, \cite{zhao2021emergent} where the value $\frac{J_{2}}{J_{2}}\sim 10^{-2}$ was observed, provided the size of the symmetry braking domains ($\xi)$ is sufficiently large. 

%Finally we would like to mention that in the case where the origin of these domains are nematic fluctuations application of stain can illuminate them, and therefore reduce the value of $J_{2}$.

{Importantly, comparing Eq. \ref{eq:J2_res} to Eq. \ref{estimate}, we find that the inhomogeneity-induced $J_2$ differs from the intrinsic one by a factor of 
\begin{equation}
\frac{J_{2}}{J^\text{int}_{2}} \sim \frac{\overline{\delta J_1^2}}{J_1^2(0)} \left(\frac{\xi}{\xi_\text{sc}}\right)^2,
\end{equation}
where we have dropped all factors of order one.}

%\section{Discussion}
%\label{sec:discussion}

{\it Discussion.--}  The vanishing of $J_1(\theta)$ as $\theta\to \pi/4$ and the existence of a higher order Josephson coupling, $J_2$ that is non-vanishing under the same conditions follow from the $d-$wave symmetry of the SC order.  There are, generically, both intrinsic and extrinsic (disorder related) contributions to $J_2(\theta)$.  However, given the extremely small c-axis superfluid density in the particular material of interest (Bi-2212), the observed~\cite{zhao2021emergent} magnitude, $|J_2(\pi/4)/J_1(0)|\sim 10^{-2}$ is notably large.  We have estimated that the intrinsic contribution is of order  $|J_2^\text{int}(\pi/4)/J_1(0)|\sim 10^{-5}-10^{-6}$ (Eq. \ref{estimate}).  What we have shown is that an extrinsic mechanism - one that derives from a fluctuating in space but random in sign first order coupling $J_1(\vec r)$ - can produce an effect of the requisite magnitude but only under rather extreme circumstances, i.e. when $\overline {\delta J_1^2}/J_1^2(0)$ is sufficiently large {\it and} the correlation length, $\xi$, is relatively long.  For instance, we can explain the observed value of $|J_2/J_1(0)|$ if we assume $\overline {\delta J_1^2}/J_1^2(0)=0.1$ and $\xi/\xi_\text{sc} \sim 10^2-10^3$.   

The existence of a non-vanishing $\delta J_1(\vec r)$ when globally $\theta=\pi/4$ 
%can be viewed as reflecting a local admixture of an s-wave component of the SC order. Thus, for the extrinsic mechanism to succeed, we need to postulate 
requires a local breaking of mirror symmetries (where the mirror plane is perpendicular to the system).  
%that is not small in magnitude, is strongly coupled to the SC order parameter, and has substantially long-range spatial correlations.  
%Large amplitude 
Twist-angle disorder leads to such symmetry breaking. However, to produce an effect of the requisite magnitude observed in Ref.~\cite{zhao2021emergent}, either the distribution of the twist angles should be a substantial fraction of $\pi/4$, or their correlation length should be larger than the sample size (about 10$\mu$m).  
% the distribution of twist angles would need to be a substantial fraction of $\pi/4$, % - something that is apparently ruled out 
% which is implausible in the devices in question~\cite{zhao2021emergent}. Alternatively, if the amplitude of the twist angle fluctuations is about $1^\circ$ (corresponding to $(\overline{\delta J_1^2})^{1/2}/J_1(0)\sim 0.02$), the correlation length $\xi$ of the angle fluctuations would need to be of the order of tens of $\mu\rm{m}$ to explain the experiments - of the order of the system size in Ref.~\cite{zhao2021emergent}. 
Some source of substantial shear-strain disorder might also be relevant, but there would need to be some reason to believe that such strain would strongly perturb the local pairing symmetry.  

We thus consider the most likely origin of the requisite disorder is %disorder pinning of some form of intrinsic electron nematic order
to be pinned domains of an otherwise intrinsic electron-nematic order~\cite{KFE}. 
There exists strong - although not universally accepted - evidence of a  tendency toward nematic order in the cuprates~\cite{ando,hinkov,bozovic,rmpkapitulnik,vojta}.  
In particular, in Bi-2212, there is direct evidence from scanning tunnelling microscopy (STM)~\cite{rmpkapitulnik,lawler,seamusvestigial,erica,chen2022identification} of local nematic order which is seen strongly and primarily at energies of order of the gap maximum (i.e. at energies at which the density of states in the superconducting state exhibits a maximum). This implies that the local nematicity has a strong effect on the local superconducting order parameter. Moreover, the fact that signatures of nematic symmetry breaking remain strong when STM features are spatially averaged over the field of view suggests a correlation length larger than the field of view, i.e. $\xi_\text{nem} > 100a$ (where $a$ is the lattice spacing).  
%In particular, 
In addition, a recent STM study~\cite{erica} on a related material (Bi-2201) also provides evidence of long-range nematic correlations.\cite{ericanote} 

There are several testable consequences of various aspects of the above line of reasoning:  1) An extrinsic effect depends strongly on $\xi$ and $\overline{\delta J_1^2}$, which could well depend on details of sample preparation. This, in turn, might explain the already mentioned fact that the twist-angle dependence is not universally observed~\cite{XueTwist}.  
% This could be accomplished artificially. 
2) The time reversal breaking phase should have spontaneous circulating currents, as shown in Fig. \ref{fig1}b. The expected order of magnitude of the in-plane magnetic moment is $m_\parallel\sim (2e/\hbar)\,(\overline{\delta J^2_1})^{1/2}\xi^3 d$. Taking $\overline{(\delta J^2_1})^{1/2} \sim 0.3\,J_1(0) \sim 1.5\cdot 10^{-5}\rm{meV}/\rm{nm}^2$,  $\xi=200\rm{nm}$ and $d=1.5\rm{nm}$, this gives $m_\parallel$ of the order of a few $\mu_B$. The associated magnetic fields should be observable near the edges of the system, and have a random sign.
3) From STM~\cite{lawler} and other~\cite{hawthorn} studies, it appears that the nematicity in Bi-2212 vanishes (or at least becomes much weaker) for hole doping concentration larger than a critical value, $p^\star \approx 0.19$.  Thus, a correspondingly strong doping dependence of the magnitude of $J_2$ would be expected if electron-nematicity plays a role in the  effect.\cite{seamus} 

From a symmetry point of view, the state of the system at $\theta=\pi/4$ is, as has been previously noted~\cite{can2021high}, a $d+id$ superconductor. 
%For a purely intrinsic case, 
In the absence of disorder (assuming that $J_2$ is generated by the intrinsic mechanism), 
this state should be fully gapped - although for numbers relevant to Bi-2212 this induced nodal gap $\Delta_\text{ind}$ would likely be immeasurably small.  
For the extrinsic case, there is an interesting question of principle whether this state is gapped or gapless. Potential disorder is expected to suppress the gap. The gap is further reduced by local Doppler shifts of the quasi-particle energies (Volovik effect~\cite{volovik1993superconductivity}) due to the presence of equilibrium currents,
 $\delta E(\vec r) =  v_F j_s(\vec r)/\kappa$. If this energy shift is larger than $\Delta_\text{ind}(\vec r)$, the gap closes.
 
Finally, we note that the present considerations are not 
%check the estimate of $j_s\sim J_1 \xi_{n}$.  In this case, we are in the gapless regime when $\xi_n/a \gg 1$.
 confined to the cuprates.  In less anisotropic systems, there may be circumstances in which the intrinsic effect dominates, and in which the resulting state at $\theta=\pi/4$ is fully gapped, as suggested in Ref. \cite{can2021high}.  It is also worth noting that very similar considerations apply to a junction between an unconventional superconductor (e.g. a d-wave superconductor) and a conventional s-wave superconductor, without need for twist angle engineering.

\section{Acknowledgments}

We are grateful to J. C. Davis, P. Kim, M. Franz, and A. Pasupathy, J. Pixley, and J. M. Tranquada for useful discussions. We thank the hospitality of the Aspen Center for Physics and the Oxford Physics Department, where part of this work was done.  SAK and AY were supported, in part, by
NSF grant No. DMR-2000987 at Stanford and SAK was further supported by a Leverhulme Trust International Professorship grant numberLIP-202-014 at Oxford.
The work BZS was funded by the Gordon and Betty Moore Foundation’s EPiQS Initiative through Grant GBMF8686.
YV and EB were supported by the European Research Council (ERC) under the European Union’s Horizon 2020 research and innovation programme (grant agreement No 817799) and by the Israel-USA Binational Science Foundation (BSF). 

% \bibliography{refs.bib}
\bibliography{main.bbl}

\appendix
\onecolumngrid

% \section{Angular dependence of $I_c$ for coherent and incoherent tunneling}
% \label{app:incoherent}

% \section{Mean-field theory}
% \label{app:MFT}
\section{Further Quantitative Considerations}
\label{app:quant}
The most crucial parameter for quantitative estimates of $J_1$ and $J_2$ is the inter-bilayer tunneling matrix element, $t_\perp$. Estimates of this parameter from band structure calculations vary from a few meV~\cite{Markiewicz2005} to as much as $20$meV~\cite{can2021high}. Moreover, $t_\perp$ may have a significant angle dependence~\cite{Song2022}. The largest of these estimates could give rise intrinsic values of $J_2/J_1$ that are compatible with the experiment of Ref. \cite{zhao2021emergent}. 

Considering the significant variation of the first principle estimates, and the fact that $t_\perp$ could exhibit a large many-body renormalization~\cite{Kumar1992}, we take a more empirical approach. Several measurable quantities are expected to be proportional to $t_\perp^2$: these include $\kappa_\perp$, the bulk c-axis critical current density $J_c$, the bulk c-axis conductivity, the critical current of the junction at $\theta=0$, and the inverse of the normal state junction resistance $R^{-1}_N$. As we enumerate below, these various quantities all give a roughly consistent estimate, $(t_\perp/t)^2 \sim 10^{-5}$ (where $t$ is the in-plane hopping matrix element). Taking $t\approx 0.25$eV, this gives $t_\perp \sim 1$meV, and the estimates of $J_2/J_1(0)$ quoted in the main text.  

As was noted in Ref.~\cite{zhao2021emergent}, the $\theta=0$ critical current density of the junction is similar to the bulk $c-$axis critical current density of optimally doped Bi-2212. This confirms that the coupling across the junction is of similar magnitude to the inter-bilayer bulk coupling. The c-axis phase stiffness inferred from the critical currents reported in Refs.~\cite{zhao2021emergent,Irie2000}, $J_c\approx 1\rm{kA}/\rm{cm}^2$, given by 
\begin{equation}
\kappa_\perp = \frac{\hbar J_c d}{2e}\approx 3.1\cdot 10^{-5}\rm{meV/nm},     
\end{equation}
is consistent within a factor of 2 with the estimate of $\kappa_\perp$ from the c-axis penetration depth, using $\lambda_c = 100\mu$m~\cite{cooper1990direct},
\begin{equation}
\kappa_\perp = \frac{1}{\mu_0 (2e/\hbar)^2}\,\frac{1}{\lambda_c^2}\approx 5.4\cdot 10^{-5} \rm{meV/nm} .     
\end{equation}
As noted in the text, these values point to a ratio $\kappa_\perp/\kappa_\parallel \approx 4\cdot 10^{-6}$. The resistivity anisotropy of optimally doped Bi-2212 at $T=T_c$ is $\rho_{ab}/\rho_c \approx 10^{-5}$~\cite{schrieffer2007handbook}, and the anisotropy tends to increase with decreasing temperature.   

A further consistency check is provided by the Ambegaokar-Baratoff relation~\cite{Ambegaokar1963}. In Ref.~\cite{zhao2021emergent}, the values of the normal state junction resistance $R_N$ and the critical current $I_c$ are reported for various samples with different twist angles. For those with $\theta$ near $0$, The product $I_c R_N$ is about $15$meV, which is smaller than (but of the order of) $\pi \Delta_{\rm{max}}/(2e)$, where $\Delta_{\rm{max}}\approx 40$meV is the maximal gap inferred from various spectroscopy experiments~\cite{Damascelli2003}. This indicates that near $\theta=0$, the critical current is not dramatically suppressed due to destructive interference (as may arise, e.g., in a dirty junction between $d$-wave superconductors). Near $\theta=\pi/4$, $I_c R_N$ is smaller by a factor of about $50$, similar to the suppression of $I_c$. 

\section{Perturbation Theory}
\label{app:pert}
%\subsection{Derivation of the current-phase relation}
Here, we provide more details on the perturbative treatment of the Hamiltonian in Eq. \ref{eq:H}. We use the following \emph{ansatz} for $\phi_n(\bm{r})$: 
\begin{equation}
\phi_{n}(\bm{r})=\begin{cases}
\,\,\,\frac{\bar{\phi}}{2}+\varphi_{n}(\bm{r}), & n>0,\\
-\frac{\bar{\phi}}{2}-\varphi_{-n+1}(\bm{r}), & n\le 0.
\end{cases}
\label{eq:ansatz}
\end{equation}
We insert Eq. \ref{eq:ansatz} into Eq. \ref{eq:H}, minimize with respect to $\varphi_n({\bm{r}})$ and linearize the resulting equations. This gives:
\begin{equation}
-\kappa\nabla^{2}\varphi_{n}+J_{\perp}\left(2\varphi_{n}-\varphi_{n+1}-\varphi_{n-1}\right)=0\,\,\,(1<n<N),\label{eq:bulk}
\end{equation}
\begin{equation}
-\kappa\nabla^{2}\varphi_{1}+J_{\perp}\left(\varphi_{1}-\varphi_{2}\right)+J_1(\bm{r})\sin\bar{\phi}=0,
\label{eq:edge1}
\end{equation}
\begin{equation}
-\kappa\nabla^{2}\varphi_{N}+J_{\perp}\left(\varphi_{N}-\varphi_{N-1}\right)=0.
\label{eq:edgeN}
\end{equation}
We insert a solution of the form
\begin{equation}
\varphi_{n}(\bm{r})=\sum_{\bm{q}}e^{i\bm{q}\cdot\bm{r}}\left(A_{\bm{q}}e^{\eta_{\bm{q}}\left(n-1\right)}+B_{\bm{q}}e^{-\eta_{\bm{q}}\left(n-1\right)}\right)\,\,\,(n>0),
\end{equation}
and $\varphi_{-n}(\bm{r})=-\varphi_{n+1}(\bm{r})$.

From Eq. (\ref{eq:bulk}),
\begin{equation}
\kappa q^{2}+2J_{\perp}\left(1-\cosh\eta_{\bm{q}}\right)=0,
\end{equation}
i.e.
\begin{equation}
\eta_{\bm{q}}=\cosh^{-1}\left(1+\frac{\kappa q^{2}}{2J_{\perp}}\right)=\ln\left[\left(1+\frac{\kappa q^{2}}{2J_{\perp}}\right)+\sqrt{\left(1+\frac{\kappa q^{2}}{2J_{\perp}}\right)^{2}-1}\right].\label{eq:cosh}
\end{equation}
The two useful limits of this expression are
\begin{equation}
\eta_{\bm{q}}\rightarrow\sqrt{\frac{\kappa}{J_{\perp}}}q,\,\,\,q\ll\sqrt{\frac{J_{\perp}}{\kappa}},
\end{equation}
\begin{equation}
\eta_{\bm{q}}\rightarrow\ln\left(\frac{\kappa q^{2}}{J_{\perp}}\right),\,\,\,q\gg\sqrt{\frac{J_{\perp}}{\kappa}}.
\end{equation}
To solve for $A_{\bm{q}}$ and $B_{\bm{q}}$, we use Eq. (\ref{eq:edgeN}), obtaining
\begin{equation}
A_{\bm{q}}=B_{\bm{q}}\alpha_{\bm{q}},
\end{equation}
where
\begin{align}
\alpha_{\bm{q}}=e^{-2\eta_{\bm{q}}\left(N-1\right)}\frac{J_{\perp}\left(e^{\eta_{\bm{q}}}-1\right)-\kappa q^{2}}{J_{\perp}\left(1-e^{-\eta_{\bm{q}}}\right)+\kappa q^{2}}.
\end{align}
Note that, in the limit $q\rightarrow 0$, $\alpha_q\rightarrow 1$, and that $\alpha_{\bm{q}}>0$ (this is shown using Eq. \ref{eq:cosh}).
For later use, we note also that (for $N>1$)
\begin{align}
\alpha_{\bm{q}} & =e^{-2\eta_{\bm{q}}\left(N-1\right)}\frac{J_{\perp}\left(e^{\eta_{\bm{q}}}-1\right)-\kappa q^{2}}{J_{\perp}\left(1-e^{-\eta_{\bm{q}}}\right)+\kappa q^{2}}\le e^{-2\eta_{\bm{q}}\left(N-1\right)}\frac{J_{\perp}\left(e^{\eta_{\bm{q}}}-1\right)}{J_{\perp}\left(1-e^{-\eta_{\bm{q}}}\right)}\nonumber \\
 & =e^{-2\eta_{\bm{q}}\left(N-1\right)}e^{\eta_{\bm{q}}}\le e^{-\eta_{\bm{q}}}.\label{eq:ineq}
\end{align}

Using Eq. (\ref{eq:edge1}), we find
\begin{equation}
B_{\bm{q}}=-\frac{J_{1}(\bm{q})\sin\bar{\phi}}{\kappa q^{2}\left(1+\alpha_{\bm{q}}\right)+J_{\perp}\left(1-e^{-\eta_{\bm{q}}}+\alpha_{\bm{q}}\left(1-e^{\eta_{\bm{q}}}\right)\right)},
\end{equation}
and hence
\begin{align}
\varphi_{1}(\bm{r}) = \frac{1}{A}\sum_{\bm{q}}e^{i\bm{q}\cdot\bm{r}}\frac{J_{1}(\bm{q})\sin\bar{\phi}\left(1+\alpha_{\bm{q}}\right)}{\kappa q^{2}\left(1+\alpha_{\bm{q}}\right)+J_{\perp}\left(1-e^{-\eta_{\bm{q}}}+\alpha_{\bm{q}}\left(1-e^{\eta_{\bm{q}}}\right)\right)}.
 \label{eq:phi1_pert}
\end{align}

At $\theta=\pi/4$ (such that $\overline{J_1}=0$), the current-phase relation of the junction is given by:
\begin{align}
I(\bar{\phi}) & =\sum_{\bm{q}}2\cos(\bar{\phi})J_1(\bm{q})\varphi(-\bm{q})=-\sum_{\bm{q}}\frac{\left|J_1(\bm{q})\right|^{2}\left(1+\alpha_{\bm{q}}\right)\sin2\bar{\phi}}{\kappa q^{2}\left(1+\alpha_{\bm{q}}\right)+J_{\perp}\left(1-e^{-\eta_{\bm{q}}}+\alpha_{\bm{q}}\left(1-e^{\eta_{\bm{q}}}\right)\right)}\nonumber \\
 & =-\sum_{\bm{q}}\frac{\left|J_1(\bm{q})\right|^{2}}{\kappa q^{2}+J_{\perp}\frac{1-e^{-\eta_{\bm{q}}}+\alpha_{\bm{q}}\left(1-e^{\eta_{\bm{q}}}\right)}{1+\alpha_{\bm{q}}}}\sin2\bar{\phi},
\end{align}
which gives Eq. \ref{eq:J2} in the main text. Note that the denominator is non-negative:
\begin{equation}
\frac{1-e^{-\eta_{\bm{q}}}+\alpha_{\bm{q}}\left(1-e^{\eta_{\bm{q}}}\right)}{1+\alpha_{\bm{q}}}=\left(1-e^{-\eta_{\bm{q}}}\right)\frac{1-\alpha_{\bm{q}}e^{\eta_{\bm{q}}}}{1+\alpha_{\bm{q}}}\ge0,
\end{equation}
where we have used Eq. (\ref{eq:ineq}). 

% \subsection{Breakdown of perturbation theory at large $\xi$}

% As mentioned in the main text, the perturbative analysis breaks down for sufficiently large $\xi$, where $\overline{\varphi^2(\bm{r})}$ becomes of order $\pi$. Then, the system breaks into domains of size $\xi$ where the phase difference is close to either $0$ or $\pi$. For $N=1$ (monolayer on either side of the junction), the size of the domain walls is of the order of $\ell \sim [\kappa/(\overline{\delta J^2})^{1/2})]^{1/2}$ (Eq. \ref{eq:l} of the main text). The crossover to the non-perturbative regime, where the phase difference deviates substantially from $\pi \pi/2$, occurs when $\xi$ exceeds $\ell$. Here, we briefly comment on the derivation of the relevant length scale in the three-dimensional case ($N\rightarrow \infty$), quoted in Eq. \ref{eq:ell_3D}. 

% We compute the variance of the local deviation of the phase from $\pi/2$ in the $N\rightarrow \infty$ case using Eq. \ref{phi1_pert}, setting $\bar{\phi}=\pi/2$ and $\alpha_{\bm{q}}=0$. Using the expression for $\eta_{\bm{q}}$ at small $q$, we obtain:
% \begin{equation}
% \overline{\varphi_{1}^{2}(\bm{r})}=\int\frac{d^{2}q}{(2\pi)^{2}}\frac{\left|J_{1}(\bm{q})\right|^{2}}{\left(\kappa q^{2}+\sqrt{\kappa J_{\perp}}q\right)^{2}}\approx\frac{\overline{\delta J_{1}^{2}}\xi^{2}}{2\pi}\int_{0}^{1/\xi}\frac{qdq}{\left(\kappa q^{2}+\sqrt{\kappa J_{\perp}}q\right)^{2}}.
% \end{equation}
% The integral is infra-red logarithmically divergent. We assume that this divergence is cut off at a small momentum scale $m$ (which can be obtained as in the )

\section{Self-consistent treatment}
We consider the Hamiltonian of Eq. \ref{eq:H} for the case $N=1$ (one layer in each side of the junction) and $\theta=\pi/4$ (i.e., $\overline{J_1}=0$). The Hamiltonian for the relative phase $\varphi(\bm{r}) = \phi_1(\bm{r}) - \phi_0(\bm{r})$ is written as: 
\begin{equation}
H=\int d^{2}r\left[\frac{\kappa}{2}\left(\nabla\varphi\right)^{2} - \delta J_1(\bm{r})\cos[ \varphi(\bm{r})])\right],   
\label{eq:H_twolayer}
\end{equation}
We now discuss the infra-red divergence encountered in the perturbative treatment, and how it can be cured by performing a self-consistent harmonic calculation. The latter analysis reveals an emergent length scale which appears in the logarithm in Eq. \ref{eq:J2_res} for $N=1$. 
A similar divergence appears, and can be treated in a similar fashion, for any finite $N$. 

The perturbative expression for the phase in the first layer is given by Eq. \ref{eq:phi1_pert} with $J_\perp=0$ and $\alpha_q=0$. Then, we find that the variance of $\varphi_1(\bm{r})$ is 
\begin{equation}
    \overline{\varphi^{2}(\bm{r})}=\int \frac{d^2 q}{(2\pi)^2}\frac{\overline{\delta J_1(\bm{q}) \delta J_1(\bm{-q})}}{\left(\kappa q^{2}\right)^{2}}. 
\end{equation}
This is infrared divergent, signalling a breakdown of the perturbative treatment at long wavelengths. To deal with this divergence, we perform a self-consistent harmonic approximation. Writing $\varphi(\bm{r}) = \pi/2 - \varphi_v (\bm{r})$, we replace Eq. \ref{eq:H_twolayer} with a variational Hamiltonian:
\begin{equation}
H_{var}=\int d^{2}r\left[\frac{\kappa}{2}\left(\nabla\varphi_{v}\right)^{2}+\frac{m^{2}}{2}\varphi_{v}^{2}-\delta J_1(\bm{r})\varphi_{v}(\bm{r})\right],
\end{equation}
and minimize $H[\varphi_{v}(\bm{r})]$ with respect to $m$, with $\varphi_{v}(\bm{r})$
taken to be the minimizer of $H_{var}$: 
\begin{equation}
\varphi_{v}(\bm{q})=\frac{\delta J_1(\bm{q})}{\kappa q^{2}+m^{2}}
\end{equation}

In real space,
\begin{equation}
\varphi_{v}(\bm{r})=\int d^{2}r'K(\bm{r}-\bm{r}')\delta J_1(\bm{r}'), \quad K(\bm{r})=\frac{1}{A}\sum_{\bm{q}}\frac{1}{\kappa q^{2}+m^{2}}e^{i\bm{q}\cdot \bm{r}}.
\end{equation}
% where
% \begin{equation}
% K(\bm{r})=\frac{1}{A}\sum_{\bm{q}}\frac{1}{\kappa q^{2}+m^{2}}e^{i\bm{q}\cdot \bm{r}}
% \end{equation}
The variational energy is given by:
\begin{equation}
\overline{H[\varphi_{v}(\bm{r})]}=A\sum_{\bm{q}}\frac{\kappa q^{2}\, \overline{\delta J_1(\bm{q}) \delta J_1(\bm{-q})}}{2\left(\kappa q^{2}+m^{2}\right)^{2}}-\int d^{2}r\,\overline{\delta J_1(\bm{r})\sin\varphi_{v}(\bm{r})},
\end{equation}
where
\begin{align}
    \int d^{2}r\,\overline{\delta J_{1}(\bm{r})\sin\varphi_{v}(\bm{r})}	&=\frac{1}{2i}\int d^{2}r\,\overline{\delta J_{1}(\bm{r})e^{i\varphi_{v}(\bm{r})}}+c.c.\nonumber\\
	&=\frac{1}{2i}\int d^{2}r\overline{\delta J_{1}(\bm{r})e^{i\int d^{2}r'K(\bm{r}-\bm{r}')\delta J_{1}(\bm{r}')}}+c.c.\nonumber\\
	&=\frac{1}{2i}\int d^{2}r\left(\frac{1}{i}\frac{\delta}{\delta\alpha(\bm{r})}\right)\overline{e^{i\int d^{2}r'\left[K(\bm{r}-\bm{r}')+\alpha(\bm{r}'))\right]\delta J_{1}(\bm{r}')}}\Big|_{\alpha=0}+c.c.\nonumber\\
	&=\frac{1}{2i}\int d^{2}r\left(\frac{1}{i}\frac{\delta}{\delta\alpha(\bm{r})}\right)e^{-\frac{1}{2}\int d^{2}r''\int d^{2}r'\left[K(\bm{r}-\bm{r}')+\alpha(\bm{r}')\right]\left[K(\bm{r}-\bm{r}'')+\alpha(\bm{r}'')\right]\overline{\delta J_{1}(\bm{r}')\delta J_{1}(\bm{r}'')}}|_{\alpha=0}+c.c.\nonumber\\
	&=\sum_{\bm{q}}\frac{\overline{\delta J_{1}(\bm{q})\delta J_{1}(-\bm{q})}}{\kappa q^{2}+m^{2}}\exp\left[-\frac{1}{A}\sum_{\bm{q}}\frac{\overline{\delta J_{1}(\bm{q})\delta J_{1}(-\bm{q})}}{\left(\kappa q^{2}+m^{2}\right)^{2}}\right],
\end{align}
% \begin{align}
% \int d^{2}r\,\overline{\delta J_{1}(\bm{r})\sin\varphi_{v}(\bm{r})}&=\frac{1}{2i}\int d^{2}r\,\overline{\delta J_{1}(\bm{r})e^{i\varphi_{v}(\bm{r})}}+c.c.\nonumber\\
% &=\frac{1}{2i}\int d^{2}r\overline{\delta J_{1}(\bm{r})e^{i\int d^{2}r'K(\bm{r}-\bm{r}')\delta J_{1}(\bm{r}')}}+c.c.\nonumber\\
% &=\frac{1}{2i}\int d^{2}r\left(\frac{1}{i}\frac{\delta}{\delta\alpha(\bm{r})}\right)\overline{e^{i\int d^{2}r'\left[K(\bm{r}-\bm{r}')+\alpha(\bm{r}'))\right]\delta J_{1}(\bm{r}')}}\Big|_{\alpha=0}+c.c.\nonumber\\
% &=\frac{1}{2i}\int d^{2}r\left(\frac{1}{i}\frac{\delta}{\delta\alpha(\bm{r})}\right)e^{-\frac{1}{2}\int d^{2}r''\int d^{2}r'\left[K(\bm{r}-\bm{r}')+\alpha(\bm{r}')\right]\left[K(\bm{r}-\bm{r}'')+\alpha(\bm{r}'')\right]\delta J^{2}f(\frac{\bm{r}'-\bm{r}''}{\xi})}|_{\alpha=0}+c.c.\nonumber\\
% &=A\int_q\frac{\overline{\delta J_{1}^{2}}\tilde{f}(\bm{q})}{\kappa q^{2}+m^{2}}\exp\left[-\int_q\frac{\overline{\delta J_{1}^{2}}\tilde{f}(\bm{q})}{\left(\kappa q^{2}+m^{2}\right)^{2}}\right],
% \end{align}
where we have taken the distribution of $J_1(\bm{r})$ to be Gaussian. 
%and $\tilde{f}(\bm{q})=\int d^{2}r\,e^{-i\bm{q}\cdot\bm{r}}f(\bm{r}/\xi)$. 
We assume that $\xi\ll \sqrt{\kappa}/m$ (we will check the validity of this assumption below). Then, $\overline{\delta J_{1}(\bm{q})\delta J_{1}(-\bm{q})}$ can be replaced by $\overline{\delta J_{1}(\bm{q}=0)^2} = A\overline{\delta J_{1}(\bm{r})^{2}}\xi^{2}$, and the upper cutoff of the momentum integrals is given by $1/\xi$. Evaluating the integrals, we obtain the variational energy
\begin{equation}
    E(m^{2})=\overline{H[\varphi_{v}]}=Am_{0}^{2}\left[\frac{1}{2}\left(\ln\frac{\kappa}{m^{2}\xi^{2}}-1\right)-\ln\frac{\kappa}{m^{2}\xi^{2}}e^{-\frac{m_{0}^{2}}{2m^{2}}}\right],
\end{equation}
where we have defined
\begin{equation}
% m_{0}^{2}=\frac{\overline{\delta J_{1}^{2}}\xi^{2}}{4\pi\kappa} 
m_{0}^{2}=\frac{\overline{\delta J_{1}^{2}}\xi^{2}}{4\pi\kappa} = \frac{\kappa}{4\pi \ell^2} \frac{\xi^2}{\ell^2}, \quad \ell^2 = \frac{\kappa}{(\overline{\delta J_{1}^{2}})^{1/2}}
\end{equation}
Where $\ell$ is the disorder length scale for $N=1$ as discussed in Eq. \ref{eq:l}.
Minimizing the energy:
\begin{equation}
\frac{\partial}{\partial m^{2}}E(m^{2})=A\frac{m_{0}^{2}}{2m^{2}}\left[-\frac{1}{2}+\left(1-\frac{m_{0}^{2}}{2m^{2}}\ln\frac{\kappa}{m^{2}\xi^{2}}\right)e^{-\frac{m_{0}^{2}}{2m^{2}}}\right]=0,    
\end{equation}
We assume that $\frac{m_{0}^{2}}{2m^{2}}\ll 1$, to be checked self-consistently. Then, the exponent $e^{-\frac{m_{0}^{2}}{2m^{2}}}\approx 1$ to leading order in $\frac{m_{0}^{2}}{2m^{2}}$, and we obtain 
\begin{equation}
\frac{m_{0}^{2}}{2m^{2}}\ln\frac{\kappa}{m^{2}\xi^{2}}\approx\frac{1}{2}.
\end{equation}
The solution, to leading order in logs, is 
\begin{equation}
% m^{2}\approx m_{0}^{2}\ln\frac{\kappa}{m_{0}^{2}\xi^{2}}=\frac{\overline{\delta J_{1}^{2}}\xi^{2}}{4\pi\kappa}\ln\frac{4\pi\kappa^{2}}{\overline{\delta J_{1}^{2}}\xi^{4}}.    
m^{2}\approx m_{0}^{2}\ln\frac{\kappa}{m_{0}^{2}\xi^{2}}=\frac{\overline{\delta J_{1}^{2}}\xi^{2}}{4\pi\kappa}\ln\frac{4\pi\kappa^{2}}{\overline{\delta J_{1}^{2}}\xi^{4}} = \frac{\kappa}{4\pi\ell^2} \frac{\xi^2}{ \ell^2} \ln \frac{4\pi \ell^4}{\xi^4}
\label{eq:msq}
\end{equation}

% Our calculation is valid if $\xi^{2}\ll\kappa/(\overline{\delta J_{1}^{2}})^{1/2}$. 
Our calculation is valid if $\xi \ll \ell$. Under this condition, the assumption 
\begin{equation}
% \frac{m_{0}^{2}}{2m^{2}}=\left(\ln\frac{4\pi\kappa^{2}}{\overline{\delta J_{1}^{2}}\xi^{4}}\right)^{-1}\ll 1  
\frac{m_{0}^{2}}{2m^{2}}= \frac{1}{2} \left(\ln \frac{4\pi \ell^4}{\xi^4}\right)^{-1}\ll 1  
\end{equation}
used above is justified. Furthermore, substituting $m$ into the minimizer $\varphi_v$, we find that
\begin{equation}
    \overline{\varphi_v (\bm{r})^2} \propto  \frac{m_0^2}{m^2}  \ll 1
\end{equation}
Where $\propto$ denotes a possible constant of order 1.
% As mentioned in the main text, in the opposite limit ($\xi^{2}\gg\kappa/(\overline{\delta J_{1}^{2}})^{1/2}$), the system breaks into domains where $\varphi$ is close to either $0$ or $\pi$; this regime is not captured in the self-consistent harmonic treatment.
As mentioned in the main text, in the opposite limit ($\xi \gg\ell$), the system breaks into domains where $\varphi$ is close to either $0$ or $\pi$; this regime is not captured in the self-consistent harmonic treatment.

In summary, we find that in the two-dimensional case and for sufficiently small $\xi$, the fluctuations of the phase difference across the junction are correlated over the emergent length scale $\sqrt{\kappa}/m \gg \ell \gg \xi$, where $\sqrt{\kappa}/m$ is given by Eq. \ref{eq:msq} and rewritten as follows.
\begin{equation}
\frac{\sqrt{\kappa}}{m} = \frac{4\pi \ell^2}{\xi} \left(\ln \frac{4\pi \ell^4}{\xi^4}\right)^{-1/2}
\end{equation}
If $\xi$ exceeds $\sqrt{\kappa}/m \gg \ell$, the system breaks into domains where the phase is locked to either $0$ or $\pi$. 

\section{Numerics}

We can test the %above 
analysis outlined in Appendices B and C by minimizing the energy numerically
for a large system for different values of $\delta J$. To this end, we use a lattice version of problem, with a Hamiltonian 
\begin{equation}
    H_{latt}=-\kappa \sum_{n=0,1,\langle ij\rangle}\cos(\phi_{i,n}-\phi_{j,n})-\sum_{i}\delta J_{i}\cos(\phi_{i,1}-\phi_{i,0}).
\end{equation}
Here, $\phi_{i,n}$ is the superconducting phase of lattice site $i$ and layer $n=0,1$, $\langle ij\rangle$ denotes a pair of nearest-neighbor sites $i$ and $j$, and $\delta J_i$ is the inter-plane coupling, taken for simplicity to be uniformly distributed in the range $[-\delta J, \delta J]$. Physically, the lattice spacing corresponds to the length scale $\xi$, over which $J_1(\bm{r})$ is correlated. 

As expected, the ground state is found to break time reversal symmetry, with the average phase difference $\phi_{i,1} - \phi_{i,0}$ close to either $-\pi/2$ or $\pi/2$. Which of the two states is selected is determined by the initial conditions of the minimization. Choosing the state with $\overline{\phi_{i,1} - \phi_{i,0}}\approx \pi/2$, We define $\varphi_i = \phi_{i,1} - \phi_{i,0} - \pi/2$, and
examine the 
correlation function
\begin{equation}
C(\bm{q}) = \overline{\varphi_{\bm{q}}\varphi_{-\bm{q}}},    
\end{equation}
where $\varphi(\bm{q}) = \sum_j e^{-i \bm{q}\cdot\bm{r_j}} \varphi_j$.
The data is fit to the form
\begin{equation}
C(q,m)=\frac{C_0}{\left(q^{2}+m^{2}\right)^{2}},\label{eq:Cfit}
\end{equation}
anticipated from the self-consistent harmonic approximation.

\begin{figure}
\includegraphics[width=0.95\textwidth]{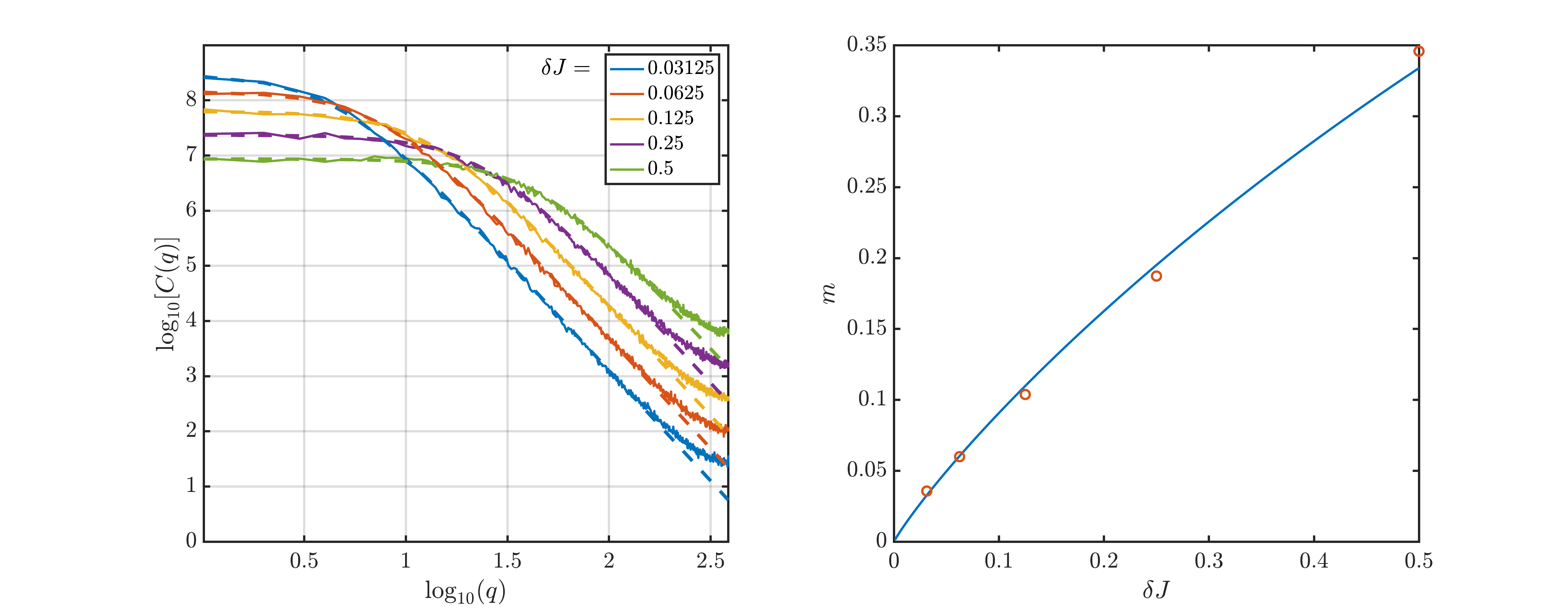}\caption{Numerical results. Left: log-log plot of the correlation function
$C(q,\delta J)=\overline{\varphi_{q}\varphi_{-q}}$. The dashed lines
are fits to Eq. \ref{eq:Cfit}. Note that at large $q$, $C(q)\sim q^{-4}$,
whereas at small $q$, $C(q)\rightarrow const.$ We use units such that $\kappa=1$. Right: the inverse
correlation length $m$ from the fits, as a function of $\delta J$.
The solid line is a fit to Eq. \ref{eq:mfit}. \label{fig:Numerical-results}}

\end{figure}

The calculated $C(q)$ (solid lines) and the fits (dashed lines) are shown in the
left panel of Fig. \ref{fig:Numerical-results} for different values of $\delta J/\kappa$. The system size is
$800\times800$, and the $C(q)$ was averaged over 8 disorder realization. We have checked that there are no significant finite size effects, even for the smallest value of $\delta J$. We find that $C_0\propto\delta J^{2}$. The right hand panel shows
the fitting parameter $m$ as a function of $\delta J$. the solid
line is a fit to Eq. (\ref{eq:msq}), of the form:
\begin{equation}
m(\delta J)=m_{1}\delta J\sqrt{\ln\frac{m_{2}}{\delta J}},\label{eq:mfit}
\end{equation}
where $m_1$ and $m_2$ are fitting parameters.

\end{document}